\documentclass[twocolumn]{revtex4-2}

\UseRawInputEncoding
\usepackage{graphicx}% Include figure files
\usepackage{dcolumn}% Align table columns on decimal point
\usepackage{bm}
\usepackage{xcolor}
\usepackage{tikz}
\usepackage{float}

\newcommand{\jb}[1]{\textcolor{black}{#1}}   % for Jonathan
\begin{document}

\title{Experimental exploration of geometric cohesion and solid fraction in columns of highly non-convex Platonic polypods}

\author{David Aponte$^{1,2}$} 
\email{da.aponte59@uniandes.edu.co} 
\email{david.aponte@umontpellier.fr}
\author{Jonathan Bar\'es$^{2}$}
\email{jonathan.bares@umontpellier.fr}
\author{Mathieu Renouf $^{2}$}
\email{mathieu.renouf@umontpellier.fr}
\author{\'Emilien Az\'ema$^{2,3,4}$} 
\email{emilien.azema@umontpellier.fr}
\author{Nicolas Estrada$^{1}$}
\email{n.estrada22@uniandes.edu.co}

\affiliation{$^{1}$Departamento de Ingeniería Civil y Ambiental, Facultad de Ingeniería, Universidad de los Andes, Bogotá, Colombia}
\affiliation{$^{2}$LMGC, Université de Montpellier, CNRS, Montpellier, France}
\affiliation{$^{3}$ Department of Civil, Geological, and Mining Engineering, Polytechnique Montréal, Montréal, Canada}
\affiliation{$^{4}$Institut Universitaire de France (IUF), Paris, France}

\date{\today}

\begin{abstract}
In this study, we investigate the stability and solid fraction of columns comprised of highly non-convex particles.
These particles are constructed by extruding arms onto the faces of Platonic solids, a configuration we term \emph{Platonic polypods}.
We explore the emergence and disappearance of solid-like behavior in the absence of adhesive forces between the particles, referred to as \emph{geometric cohesion}.
This investigation is conducted by varying the number of arms of the particles and the thickness of these arms.
To accomplish this, columns are assembled by depositing particles within a cylindrical container, followed by the removal of the container to evaluate the stability of the resulting structures.
Experiments were carried out using three distinct materials to assess the influence of the friction coefficient between the grains.
Our findings reveal that certain granular systems exhibit geometric cohesion, depending on their geometrical and contact properties.
Furthermore, we analyze the initial solid fraction of the columns, demonstrating that these arrangements can achieve stability even at highly loose states, which contrasts with traditional granular materials.
\end{abstract}

\maketitle

\section{Introduction}

The shape of particles in granular materials has long been a focal point of research.
Recent years have seen an abundance of studies aimed at understanding the influence of angularity, elongation, and platyness on the geometrical and mechanical properties of granular materials.
Particularly noteworthy is the flourishing interest in granular materials composed of particles with highly non-convex shapes, representing an extreme in particle morphology \cite{cantor2022_pp}.
For some recent examples, see Refs. \cite{Stannarius2022, Conzelmann2022, Karapiperis2022, Savoie2023, Tran2024, Aponte2024}.
These materials have gained significant attention due to their remarkable behaviors, including peculiar flowing properties \cite{Mohammadi2022,wang2024_arx}, exceptionally low solid fractions \cite{Olmedilla2018, Conzelmann2022, Tran2024}, and solid-like behavior arising solely from the geometric attributes of the grains and the friction between them.
This phenomenon, termed \emph{geometric cohesion}, was first pointed out by S.V. Franklin in 2012 \cite{Franklin2012}.

Geometric cohesion has been observed in granular materials composed of bars \cite{Phillipse1996, Stokely2003, Blouwolff2006, Desmond2006, Trepanier2010, Jung2023, Heussinger2023}, staples \cite{Franklin2012, Gravish2012, Franklin2014, Marschall2015, Karapiperis2022, Savoie2023}, Z-shaped particles \cite{Murphy2016, Karapiperis2022}, star-shaped particles \cite{Malinouskaya2009, deGraaf2011, Athanassiadis2014, Aponte2024}, and polypods \cite{deGraaf2011, Ludewig2012, Zhao2016, Bares2017, Zhao2017, Olmedilla2018, Zhao2020, Conzelmann2022, Tran2024}.
However, a comprehensive analysis of this property, systematically identifying the characteristics that enable its emergence or lead to its disappearance, is lacking in many of these studies.
This knowledge gap poses challenges in interpreting new findings and hinders the practical utilization of these materials.

The objective of this study was to contribute to the systematic investigation of granular materials exhibiting geometric cohesion.
To achieve this goal, we conducted over $3\,000$ experiments utilizing granular materials comprised of highly non-convex particles, which were fabricated by extruding arms onto the faces of Platonic solids, termed \emph{Platonic polypods}.
We systematically varied the number and thickness of these arms, as well as the coefficient of friction, $\mu$,  between the particles, employing three distinct materials: High Density Polyethylene (HDPE, $\mu \approx 0.2$), Ethylene Propylene Diene Monomer (EPDM, $\mu \approx 0.5$), and Polyamide 12 (PA12, $\mu \approx 0.8$).
Specifically, we utilized HDPE and EPDM to produce hexapods with varied arm thicknesses relative to the particle size, while PA12 was employed for 3D-printing polypods with varying numbers of arms, ranging from $4$ to $20$. 
Subsequently, these particles were assembled into granular columns of three different diameters ($D \in [33,62,97]$~mm) and subjected to stability testing.

Through this experimental setup, we were able to observe the emergence and disappearance of geometric cohesion, while also quantifying a key microstructural parameter, the solid fraction, in the initial state of the arrangements.

Firstly, our findings indicate that the granular materials under study can indeed demonstrate geometric cohesion, with the manifestation of this phenomenon depending upon factors such as the number and thickness of arms, as well as the friction coefficient between particles.
Secondly, our investigation reveals that granular systems constructed with Platonic polypods exhibit remarkably low solid fractions in comparison to conventional granular materials.
These findings not only enrich the existing experimental knowledge base but also validate previous observations obtained by means of numerical simulations.

This article is structured as follows:
Section \ref{Methods} provides an overview of the methodology employed, encompassing the fabrication of grains and the experimental setup.
Section \ref{Results} details our findings concerning the stability of the columns and their solid fraction in the initial state.
Finally, Section \ref{Conclusions} summarizes the key findings and presents our conclusions.

\section{Methods}
\label{Methods}

\subsection{Materials and particles' construction}
\label{Sec_Particle_materials_and_construction}

To investigate the influence of the friction coefficient at particle contacts, we utilized three distinct materials for particle fabrication.

\begin{figure}
\centering
\includegraphics[width=0.9\columnwidth]{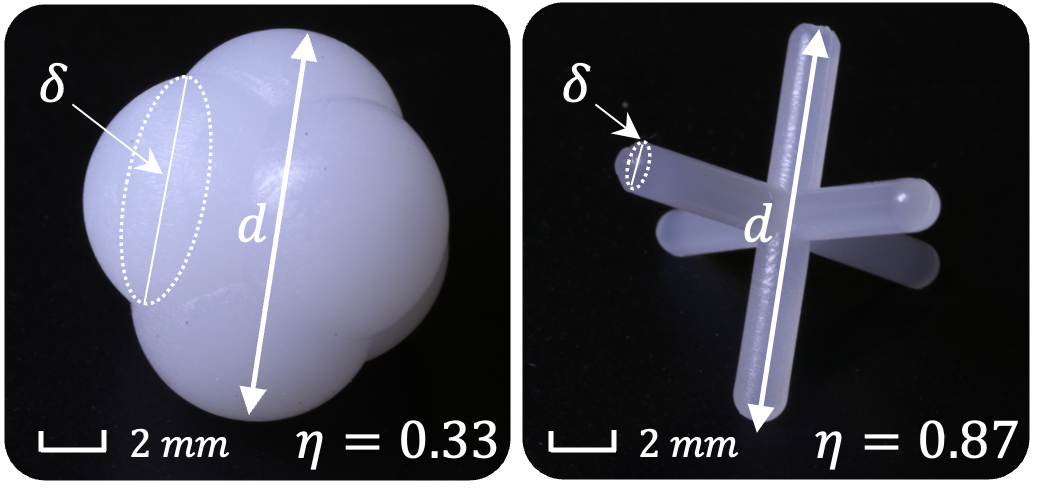}(a)
\includegraphics[width=0.9\columnwidth]{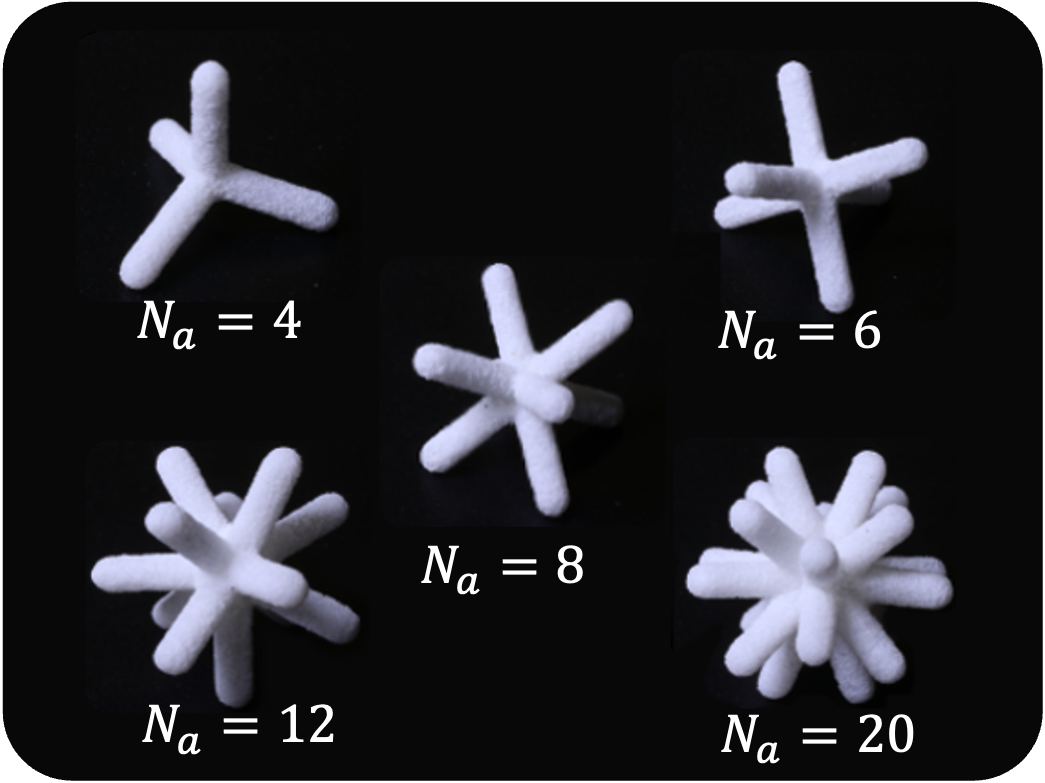}(b)
\caption{Particle description:
All particles had a size (i.e., the diameter of the circumsphere of the polypod) of $d=1.2$ cm.
(a) Particles were constructed using HDPE and EPDM, with varying arm thickness $\delta$.
This property was quantified using the aspect ratio $\eta$ of the arms.
The examples shown in this figure represent the lowest and highest $\eta$ values utilized in our experiments.
(b) Using PA12, we fabricated polypods with different numbers of arms $N_a\in[4, 6, 8, 12, 20]$, all with $\eta=0.87$.
}
\label{Fig_1_types_of_particles}
\end{figure}

The first material used was High Density Polyethylene (HDPE), which was liquefied and injected into metal molds to form particles of specific shapes.
All particles were hexapods, with varying arm thickness relative to the particle size.
To quantify the aspect ratio of the arms, that we also call \jb{\emph{concavity}}, we introduced the dimensionless parameter $\eta$, defined as:
\begin{equation}
\eta=\frac{(d-\delta)}{d},
\end{equation}
where $d$ represents the diameter of the particle and $\delta$ denotes the thickness of the arms.
The particles were constructed with a standardized size of $d=1.2$ cm, and $\eta$ values ranged from 0.33 to 0.87.
Figure \ref{Fig_1_types_of_particles}(a) shows two particles fabricated from HDPE, displaying $\eta$ values of 0.33 and 0.87, with $d$ and $\eta$ indicated for each particle.
The interparticle friction coefficient $\mu$ was measured in the lab, finding it to be close to 0.3.

The second material utilized in this study was Ethylene Propylene Diene Monomer (EPDM).
Similar to the process employed for HDPE, EPDM particles were formed by injection moulding.
The aspect ratio and number of arms were consistent with those used for HDPE particles.
From experimental measurements, we determined the friction coefficient $\mu \approx 0.5$.
Consequently, the primary distinction between HDPE and EPDM materials lies in their respective friction coefficients between particles.
Although the stiffness of these two materials differs, this property does not significantly influence our experiments, as the stresses induced by the self-weight of the columns remain low.

To explore particles with varying numbers of arms $N_a$, we employed a third material, Polyamide 12 (PA12).
Utilizing 3D printing Slice Laser Sintering (SLS) method, particles were fabricated with $N_a$ values ranging from $4$ to $20$, corresponding to the number of faces of Platonic polyhedra.
From experimental measurements, we determined $\mu \approx 0.8$.
All particles constructed using PA12 had the same aspect ratio, $\eta=0.87$.
Figure \ref{Fig_1_types_of_particles}(b) shows the five distinct types of particles fabricated with PA12.

\subsection{Experimental setup}
\label{Sec_experimental_setup}

To assess the stability of the arrangements formed with the particles described in the preceding subsection, we performed a series of column collapse tests.
These experiments have been demonstrated as effective as in previous studies \cite{Zhao2016,Aponte2024}.
Initially, the two halves of a plexiglass cylinder were positioned together on a flat, rough surface, covered with sandpaper with a grit size of $40$.
Subsequently, a predetermined quantity of particles was carefully poured into the cylinder.
Following this, the two halves of the cylinder were slowly separated.
The distribution of experiments conducted for each material and type of particles is summarized in Table \ref{tab_experiments}.
Consistently with prior literature \cite{Zhao2016,Aponte2024}, a substantial number of repetitions were performed to ensure the statistical significance of the results.

\begin{table}
\centering
\caption{
Number of tests carried out for each material, aspect ratio $\eta$, number of arms $N_a$, container size $D/d$, and range in the number of particles $N_p$.
The sum shows that more than 3000 experiments were conducted.
}
\begin{tabular}[t]{c|c|c|c|c}
\hline
\hline
\textbf{Material} & $\bm{\eta}$ ($\bm{N_a}$) & $\bm{D/d}$ & $\bm{N_p}$ & \textbf{No. of tests}\\
\hline
     & 0.33 (6) &.            & 23 ... 282  & 87\\
     & 0.50 (6) &             & 25 ... 328  & 107\\
     & 0.58 (6) &             & 25 ... 359  & 129\\
     & 0.67 (6) &             & 27 ... 374  & 123\\
HDPE & 0.71 (6) & 2.75, 5.16, & 29 ... 400  & 110\\
     & 0.75 (6) & and 8.08    & 32 ... 431  & 95\\
     & 0.79 (6) &             & 36 ... 487  & 109\\
     & 0.83 (6) &             & 40 ... 544  & 129\\
     & 0.88 (6) &             & 49 ... 667  & 80\\
\hline
     & 0.33 (6) &.            & 89 ... 364  & 77\\
     & 0.50 (6) &             & 119 ... 436 & 90\\
     & 0.58 (6) &             & 27 ... 472  & 83\\
     & 0.67 (6) &             & 31 ... 461  & 86\\
EPDM & 0.71 (6) & 2.75, 5.16, & 33 ... 554  & 91\\
     & 0.75 (6) & and 8.08    & 36 ... 615  & 108\\
     & 0.79 (6) &             & 38 ... 590  & 146\\
     & 0.83 (6) &             & 49 ... 574  & 150\\
     & 0.88 (6) &             & 62 ... 718  & 194\\
\hline
     & 0.88 (4) &.            & 82 ... 949  & 260\\
     & 0.88 (6) &             & 88 ... 1\,231 & 213\\
PA12 & 0.88 (8) & 2.75, 5.16, & 78 ... 1\,282 & 204\\
     & 0.88 (12)& and 8.08    & 98 ... 1\,282 & 208\\
     & 0.88 (20)&             & 85 ... 1\,436 & 182\\
\hline
     &          &             &             & \textbf{3\,061}\\  
\hline
\hline
\end{tabular}
\label{tab_experiments}
\end{table}

After the removal of the cylinder, two distinct behaviors were observed:
(i) In certain experiments, the material collapsed, resulting in the formation of a pile with an approximately conical shape.
For the purposes of this paper, we will refer to this behavior as a ``frictional response".
(ii) Conversely, in other experiments, the arrangement retained its initial shape, forming a free-standing column and exhibiting characteristics akin to a solid.
This behavior will be termed a ``cohesive response" throughout the paper.

\begin{figure}
\centering
\includegraphics[width=0.9\columnwidth]{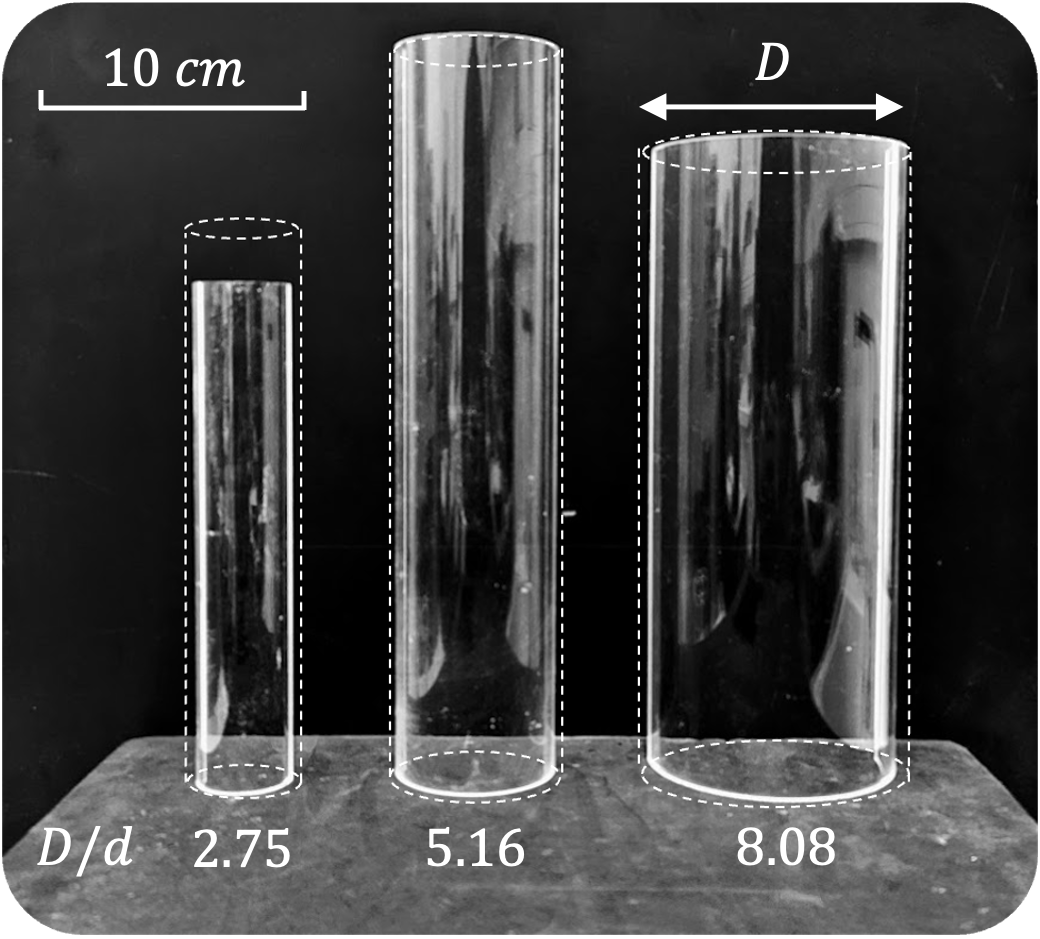}(a)
\includegraphics[width=0.9\columnwidth]{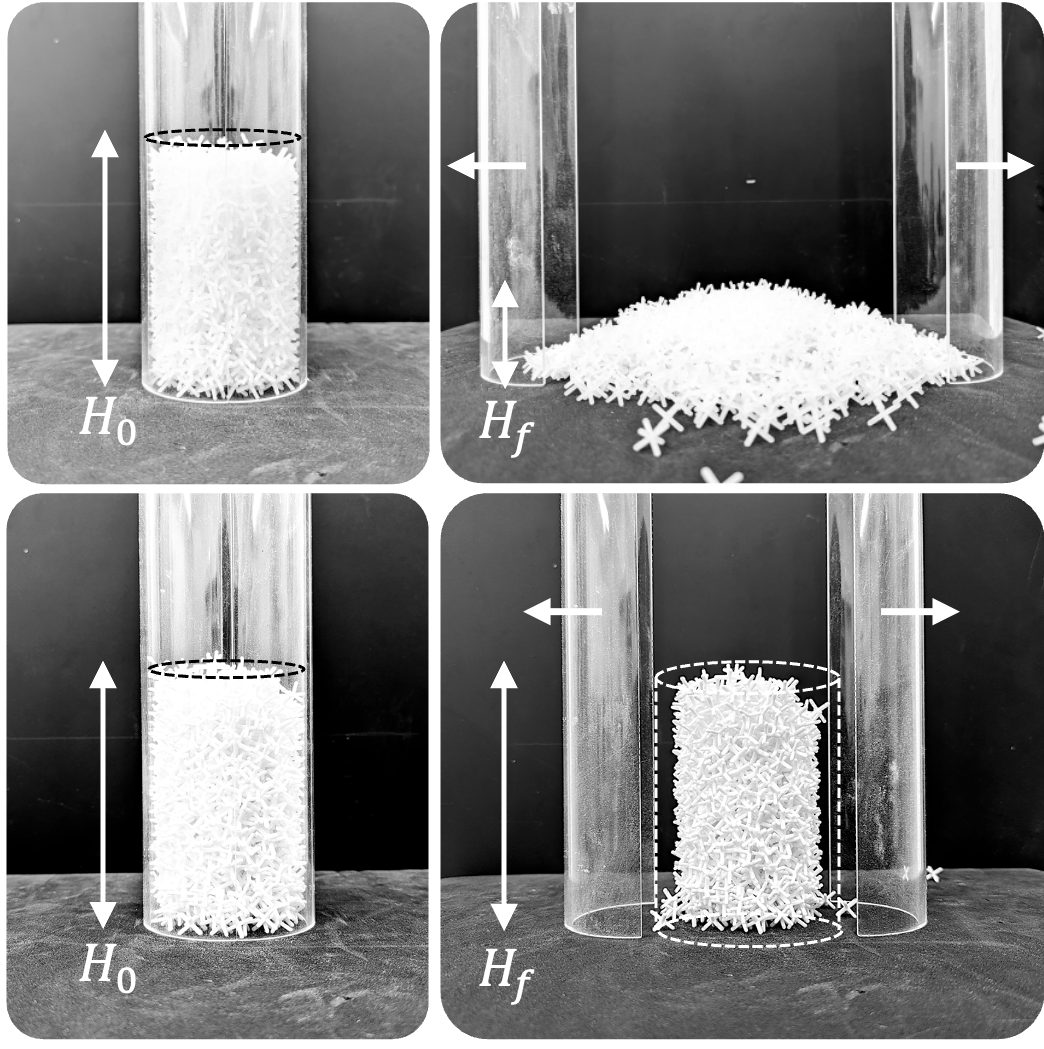}(b)
\caption{Experimental setup:
(a) For the construction of the columns, we employed three cylinders with varying internal diameters $D$.
(b) Initial (left) and final (right) states of the experiment.
Top: A column composed of hexapods built with HDPE and an aspect ratio $\eta=0.87$, showing a frictional response.
Bottom: A column composed of hexapods built with PA12 and $\eta=0.87$, showing a cohesive response.
The initial and final column heights, $H_0$ and $H_f$ respectively.
}
\label{Fig_2_Experimental_setup}
\end{figure}

To examine the influence of column size, three cylinders were utilized, with diameters $D$ measuring $2.75$, $5.16$, and $8.08$ times the size of the particles $d=12$ mm.
Figure \ref{Fig_2_Experimental_setup}(a) shows the three cylinders employed, while Fig. \ref{Fig_2_Experimental_setup}(b) shows the initial and final states of the experiment, presenting both frictional and cohesive responses.
Additionally, graphical representations of the initial and final heights of the column, denoted as $H_0$ and $H_f$ respectively, are provided.

To quantitatively assess the stability of the columns, we employed the collapse ratio $r$, as already defined in Ref. \cite{Zhao2016}, given by
\begin{equation}
r=\frac{N_{\rm{out}}}{N_p},
\label{Eq_r}
\end{equation}
where $N_{\rm{out}}$ represents the number of particles located outside the initial column volume, and $N_p$ denotes the total number of particles used in the experiment.
For columns exhibiting a cohesive response, $r$ is anticipated to approach zero, while for those demonstrating a frictional response, $r$ tends to be significantly higher, close to one.
Furthermore, to assess the degree of compactness of the samples, we calculated the initial solid fraction $\phi$:
\begin{equation}
\phi=\frac{V_p}{V},
\label{Eq_phi}
\end{equation}
where $V_p$ denotes the volume occupied by the particles, and $V$ represents the total volume.

\section{Results}
\label{Results}

\subsection{Columns' stability}

\begin{figure*}
\centering
\includegraphics[width=2\columnwidth]{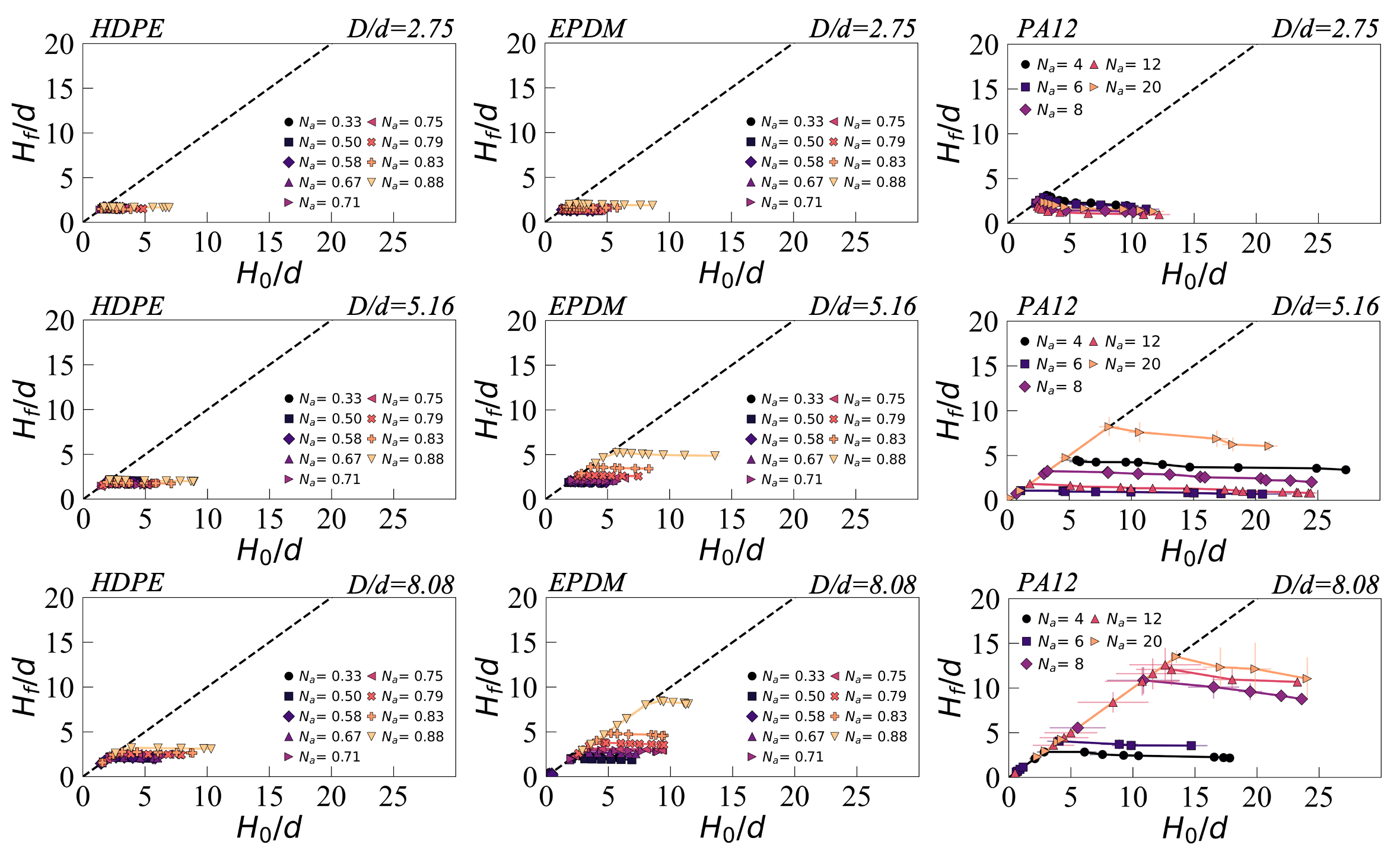}
\caption{Final height $H_f$ as a function of the initial height $H_0$ of the columns, both \jb{are} normalized by the particle size $d$.
Rows correspond to the container diameter $D$, while the columns correspond to the materials used (i.e., HDPE, EPDM, and PA12).
For HDPE and EPDM, the series represent different aspect ratios $\eta$ of the arms; all particles for these materials were hexapods ($N_a=6$).
For PA12, the series represent polypods with varying numbers of arms $N_a$, all with an aspect ratio $\eta=0.87$.
For each series, the data points correspond to columns built with different numbers of particles $N_p$ and, consequently, with different values of $H_0$.
The error bars represent the standard deviation of sets of experiments with the same $N_p$.
}
\label{Fig_H0vsHf}
\end{figure*}

Figure \ref{Fig_H0vsHf} shows the relationship between the initial height $H_0$ and the final height $H_f$ of the columns, both normalized by the particle size $d$.
The rows correspond to different container diameters $D/d$ (specifically, $2.75$, $5.16$, and $8.08$), while the columns represent the three materials employed: HDPE, EPDM, and PA12 (refer to Sec. \ref{Sec_Particle_materials_and_construction} for definitions).
For HDPE and EPDM, the series represent particles with varying aspect ratios $\eta$ of the arms, all of which were hexapods ($N_a=6$).
Conversely, for PA12, the series represent polypods with different $N_a$, each with an aspect ratio $\eta=0.87$.
Each series depicts what we term a ``system", comprising experiments with columns featuring varying numbers of particles $N_p$ and consequently different initial heights $H_0$.
This approach facilitated the examination of column behavior relative to size, thus elucidating the transition from a frictional to a cohesive response.
%NE: I'm not fully satisfied with the use of the term "system" in this sense. Do you have alternative ideas? This is important, since we use this term a lot in the following paragraphs ...
%JB: Why not to keep series...

\begin{figure*}
\centering
\includegraphics[width=2\columnwidth]{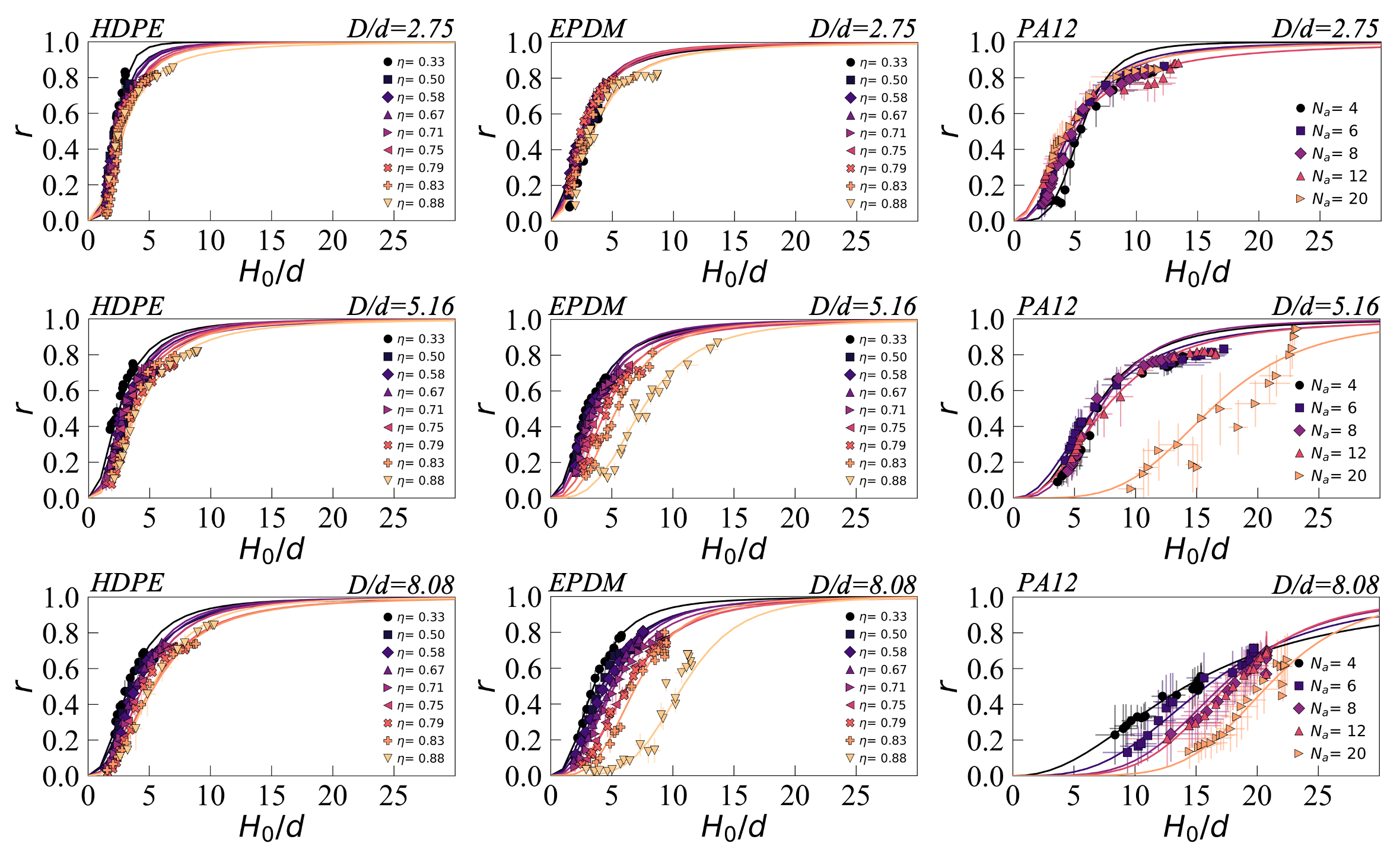}
\caption{
Collapse ratio $r$ as a function of the normalized initial height $H_0/d$.
The rows correspond to the container diameter $D$, while the columns correspond to the  materials used.
For HDPE and EPDM, the series represent different aspect ratios $\eta$ of the arms; all particles for these materials were hexapods ($N_a=6$).
For PA12, the series represent polypods with varying numbers of arms $N_a$, all with an aspect ratio $\eta=0.87$.
For each series, the data points correspond to columns built with different numbers of particles $N_p$ and, consequently, with different values of $H_0$.
The error bars represent the standard deviation of sets of experiments with the same $N_p$.
The lines represent the best fit for Eq. \ref{Eq_r}.
}
\label{Fig_rvsH0d}
\end{figure*}

In the first row (i.e., for the thinnest cylinder), it is evident that $H_f/d$ consistently remains smaller than $H_0/d$.
This observation indicates that all systems exhibited a frictional response, characterized by column collapse followed by the formation of a heap.
This pattern is similarly observed in the first column, corresponding to HDPE, which possesses the lowest friction coefficient $\mu$.
However, as $D/d$ and $\mu$ increase, certain systems stand out.
Specifically, for a range of column sizes, $H_f/d$ approximately equals $H_0/d$ (e.g., see rows-columns: 2-2, 2-3, 3-2, and 3-3).
This suggests that these columns exhibited stability and displayed a solid-like behavior, indicative of geometric cohesion, once the container was removed.
Notably, for EPDM, this behavior is observed primarily for particles with the largest $\eta$ (i.e., $\eta=0.87$).
Similarly, for PA12, this behavior is evident for particles with $N_a$ values of $8$, $12$, and $20$, with the most cohesive system observed for $N_a=20$.
For these cohesive systems, it is noteworthy that beyond a certain column size, $H_f$ progressively deviates from $H_0$, signifying the onset of a gradual transition from cohesive to frictional response, as elaborated in detail in \cite{Aponte2024}.
In summary, these results highlight that, besides particle shape, both increasing column width and enhancing the friction coefficient between particles promote the emergence of geometric cohesion.
Larger cylinder diameters enhance particle-base contact, thereby preventing column toppling, while higher friction coefficients enhance the stability of interlocked interactions, a local mechanism identified as crucial for geometric cohesion in previous studies \cite{Aponte2024}.

\subsection{Collapse ratio}

While the relationship between final and initial heights, $H_f$ and $H_0$ respectively, serves to identify and differentiate frictional and cohesive responses, this representation may not adequately capture the crossover between both regimes.
To better elucidate this transition, it is more informative to characterize our experiments in terms of the collapse ratio $r$, as defined in Eq. \ref{Eq_r}.
Figure \ref{Fig_rvsH0d} shows $r$ as a function of $H_0/d$ for the three container sizes and three materials.
For systems displaying a frictional response (e.g., all systems in the first row and first column of the figure), $r$ exhibits rapid growth with increasing $H_0/d$, indicating that even the shortest columns collapse, eventually approaching unity as column height increases.
However, as shown in Fig. \ref{Fig_H0vsHf}, a distinct behavior is observed in certain systems within larger containers, thinner arms, and more frictional materials.
These systems demonstrate geometric cohesion, transitioning from a frictional to a cohesive response.
For shorter columns, $r$ remains small, sometimes approaching zero, but beyond a critical column height, it gradually increases before asymptotically approaching unity, mirroring the behavior of systems with a frictional response.
This transition is notably observed in EPDM particles with \jb{concavity} $\eta=0.87$, as well as in PA12 particles with $N_a$ values of $8$, $12$, and $20$.
Once again, the emergence of geometric cohesion is facilitated by wider containers and higher local friction coefficients.

\begin{figure}
\centering
\includegraphics[width=0.9\columnwidth]{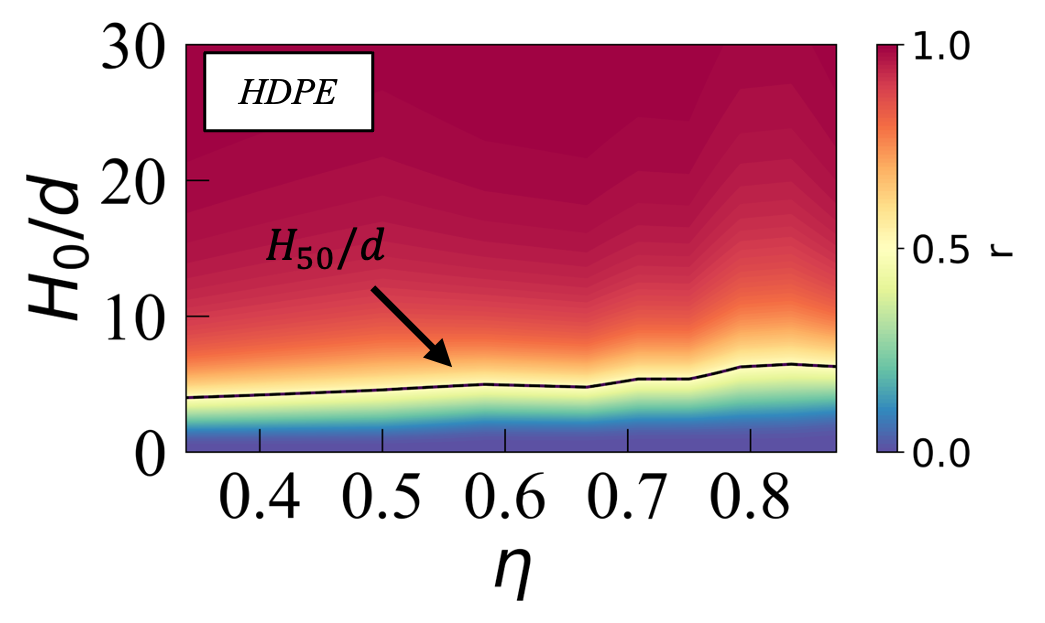}(a)
\includegraphics[width=0.9\columnwidth]{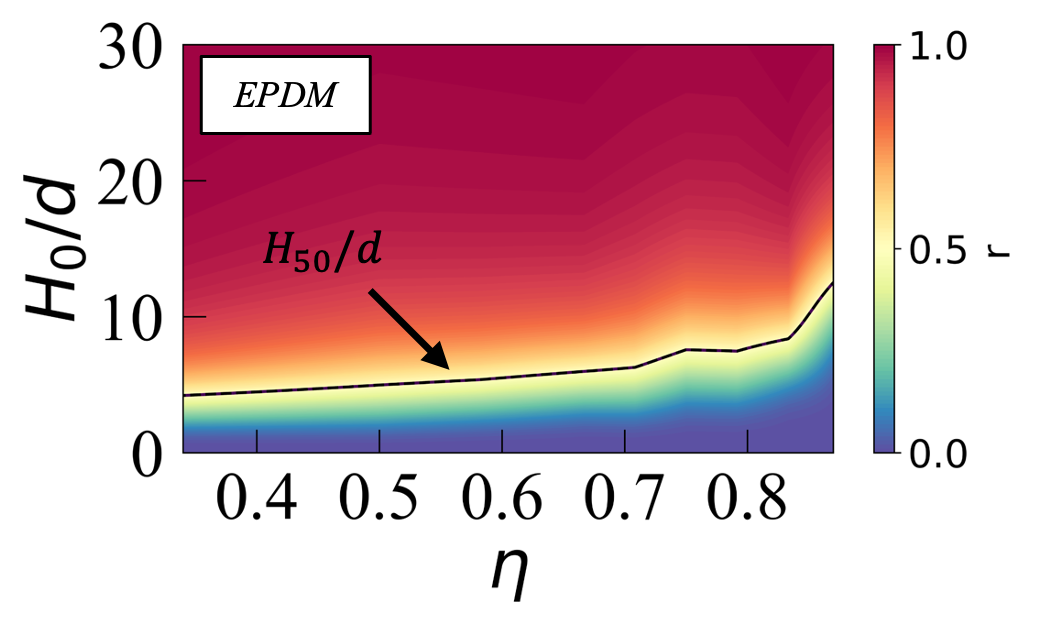}(b)
\includegraphics[width=0.9\columnwidth]{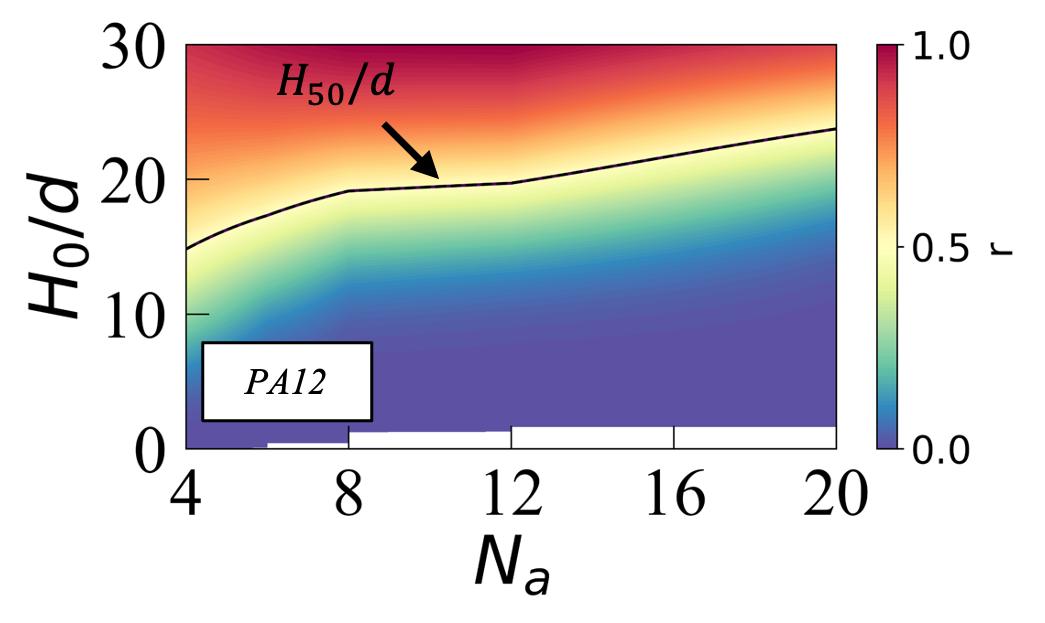}(c)
\caption{
Color maps showing the evolution of the collapse ratio $r$ of the columns as a function of the normalized initial height $H_0/d$, for columns composed of Platonic polypods with varying aspect ratio $\eta$ ((a) for HDPE and (b) for EPDM) and with varying number of arms $N_a$ for PA12 (c).
The color scale is proportional to $r$.
The evolution of the crossover height $H_{50}$ (see text for a definition) is shown in the three figures using a black line.
}
\label{Fig_H50}
\end{figure}

To quantify the system size that characterizes the transition between a frictional and a cohesive response, we employ a function introduced in \cite{Zhao2016} to describe the evolution of $r$:
\begin{equation}
\label{Eq_r}
r=\frac{1}{(H_{50}/H_0)^{\alpha}+1},
\end{equation}
where $H_{50}$ represents the height for $r=0.5$, and $\alpha$ is a fitting parameter.
Notably, $H_{50}$ can be interpreted as the length scale characterizing the transition from a frictional to a cohesive response.
Figure \ref{Fig_H50} shows color maps of $r$ as functions of (a) $\eta$ for HDPE, (b) $\eta$ for EPDM, and (c) $N_a$ for PA12, with the evolution of the crossover height $H_{50}$ plotted with a black line.
These maps can be seen as stability diagrams of the granular columns.
Firstly, from Figs. \ref{Fig_H50}(a) and (b) it is observed that $H_{50}$ increases with $\eta$, and that significant values of $H_{50}$ are predominantly observed for particles with $\eta=0.87$.
Secondly, from Fig. \ref{Fig_H50}(c) it is observed that $H_{50}$ increases with $N_a$, with the largest values observed for $N_a=20$.
The large values of $H_{50}$ shown in Fig. \ref{Fig_H50}(c) also show that the stability of the columns is strongly enhanced by using large friction coefficients $\mu$.
This is consistent with the trends observed in Figs. \ref{Fig_H0vsHf} and \ref{Fig_rvsH0d}.
These findings underscore the crucial interplay between particle shape and contact friction, both essential conditions for the emergence of geometric cohesion.

\subsection{Solid fraction}

\begin{figure}
\centering
\includegraphics[width=0.9\columnwidth]{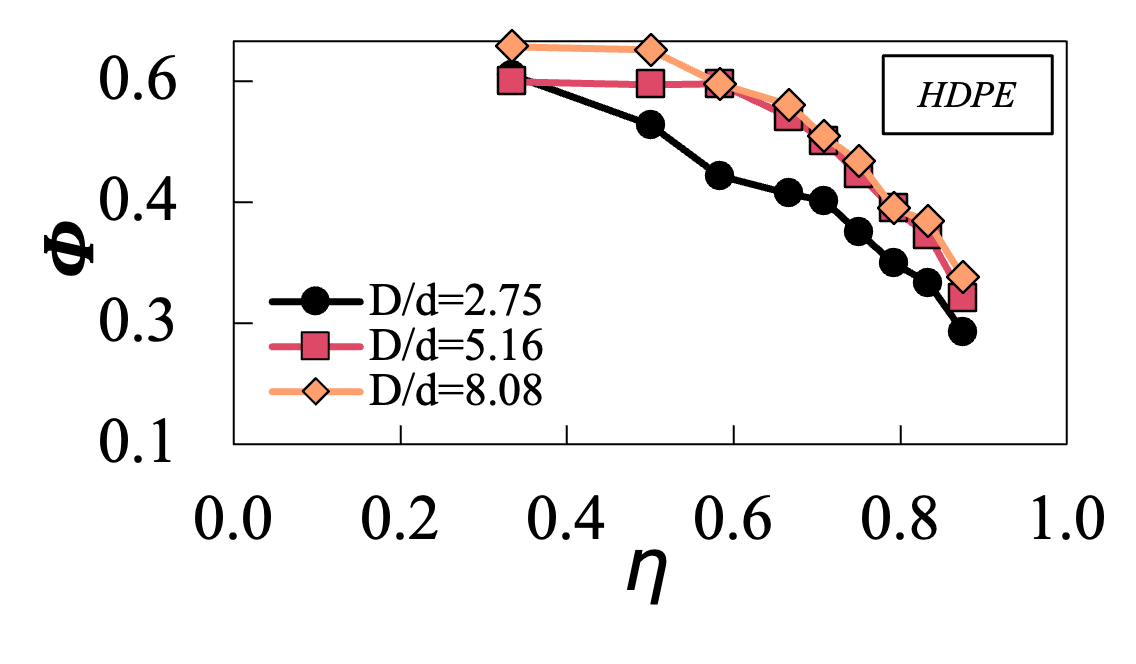}(a)
\includegraphics[width=0.9\columnwidth]{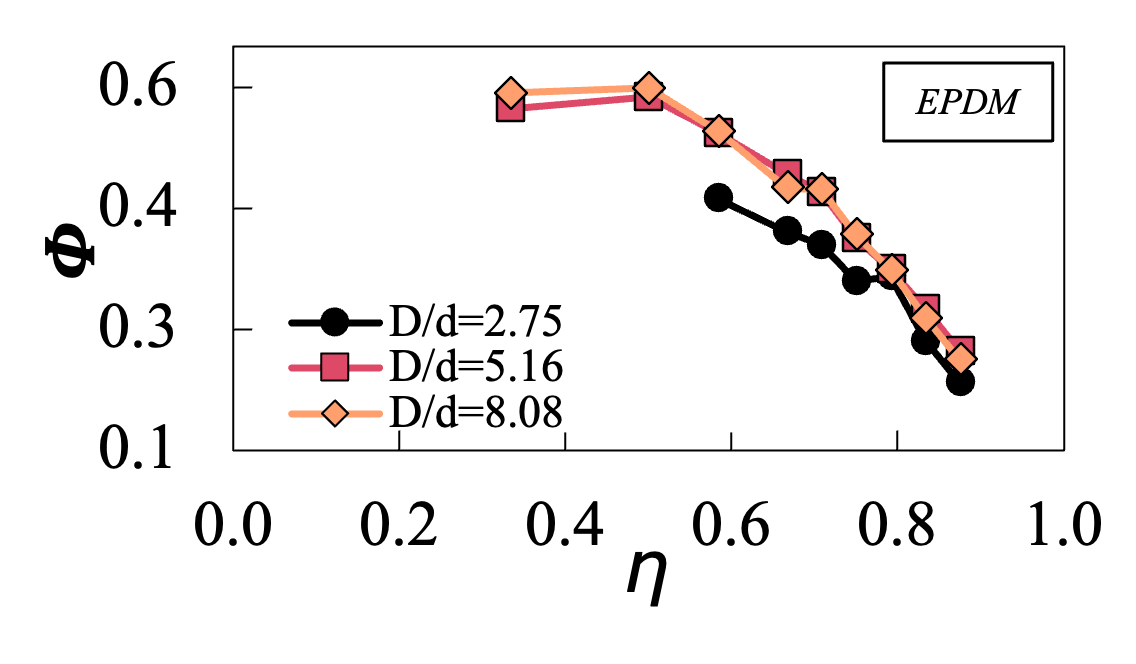}(b)
\includegraphics[width=0.9\columnwidth]{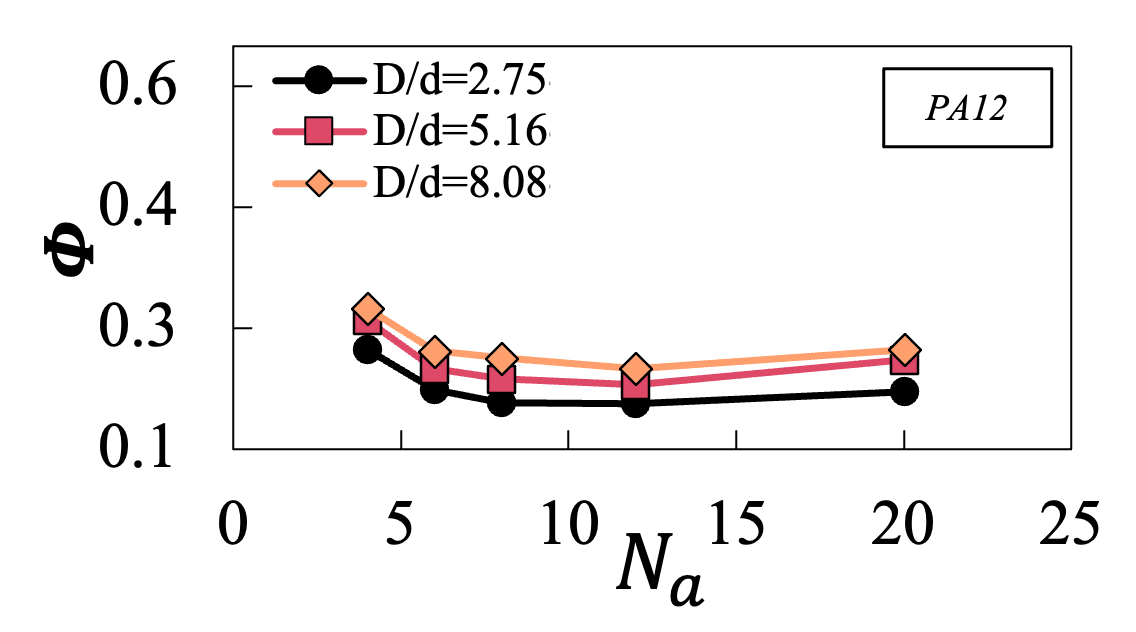}(c)
\caption{Packing fraction $\phi$, as a function of: (a) the aspect ratio $\eta$ of the arms for HDPE material, (b) $\eta$ for EPDM material, and (c) the number of arms $N_a$ for PA12 material.
For each series, the data points correspond to the mean calculated from each system, including columns built with different numbers of particles $N_p$ and, consequently, with different values of $H_0$.
The colors represent the normalized container size $D/d$.
}
\label{Fig_Phi}
\end{figure}

Figure \ref{Fig_Phi} shows the solid fraction $\phi$ of the systems in their initial state (i.e., before removing the container) as a function of: (a) the aspect ratio $\eta$ of the arms for HDPE, (b) $\eta$ for EPDM, and (c) the number of arms $N_a$ for PA12, for the three cylinder sizes $D/d$.
Firstly, from Figs. \ref{Fig_Phi}(a) and (b), it is evident that $\phi$ decreases with increasing $\eta$, transitioning from values commonly observed in typical 3D granular materials, approximately $0.6$, to significantly lower values, approaching $0.2$ (i.e., only 20\% of the volume occupied by the particles!).
Secondly, from Fig. \ref{Fig_Phi}(c), it can be seen that the evolution of $\phi$ with $N_a$ is non-monotonic, with the solid fraction appearing to be minimal for $N_a=12$.
This observation is understandable, as the addition of arms to a polypod eventually approaches a spherical shape; consequently, as $N_a$ increases, two opposing phenomena occur: the volume of the polypod increases, decreasing $\phi$, but the intercenter distance increases, further reducing $\phi$.
Thirdly, by comparing the values shown in Figs. \ref{Fig_Phi}(a) and (b) with those presented in Fig. \ref{Fig_Phi}(c), it can be deduced that $\phi$ exhibits a significant decrease with increasing the friction coefficient $\mu$.
Lastly, thin containers contribute to diminishing $\phi$, as the relative influence of the walls becomes more pronounced.
These results underscore the considerable challenge in compacting strongly \jb{concave} particles with high friction coefficients into dense arrangements.

\section{Conclusions}
\label{Conclusions}

In this investigation, we conducted a comprehensive series of experiments using columns composed of highly non-convex particles.
These particles, termed Platonic polypods, were constructed by injection moulding and utilizing a 3D SLS printer to extrude arms onto the faces of the five Platonic solids.
Three different materials were employed, allowing us to investigate the effects of the number of arms, arm thickness, and coefficient of friction between particles.
Particularly, we aimed to ascertain whether these systems demonstrate solid-like behavior solely due to particle shape and contact friction, in the absence of adhesive forces between particles, a phenomenon referred to as geometric cohesion by several authors, initially proposed by S.V. Franklin in 2012 \cite{Franklin2012}.
Additionally, we examined a first-order micromechanical parameter of these arrangements, their solid fraction, in the initial state of the experiments.

First, we demonstrate that some of the studied granular materials exhibit geometric cohesion, depending upon the number of arms and their thickness.
Notably, the materials that exhibit this property are those constructed with Platonic polypods featuring a high number of thin arms and a sufficiently large friction coefficient.
In our experiments, the particle shape that produces the most cohesive response is derived from an icosahedron (i.e., a polypod with 20 arms) and employing a highly frictional material (e.g., PA12).
We show that the manifestation of geometric cohesion is strongly dependent upon column size.
Small columns behave akin to solids, whereas an incremental column size prompts a gradual transition from cohesive to frictional behavior, akin to conventional granular materials.
Naturally, the size threshold at which this crossover occurs relies on particle shape and frictional attributes.
These findings conceptually align with our two-dimensional simulations \cite{Aponte2024}.
Additionally, we demonstrate that for geometric cohesion to emerge, column bases must surpass a certain size relative to particle dimensions; otherwise, columns are prone to collapse via toppling.
This observation was previously presented by Zhao et al. in 2016 \cite{Zhao2016} through experiments employing hexapods.

Secondly, we demonstrate that the solid fraction of the arrangements diminishes as the thickness of the arms decreases, while observing a non-monotonic relationship between the solid fraction and the number of arms of the polypods.
Among our experiments, columns constructed with polypods featuring $12$ arms from PA12 exhibit the lowest solid fraction.
The solid fraction in these arrangements can be as low as $0.2$, indicating that these granular materials can display solid-like behavior even when $80$\% of the space is empty.
Recently, similar findings were reported by by Olmedilla et al. via experiments and simulations involving equiaxed dendrites \cite{Olmedilla2018}, by Conzelmann et al. using experiments and simulations with various artificial particle shapes \cite{Conzelmann2022}, and by Tran et al. through Discrete Element Method simulations of hexapods \cite{Tran2024}.

The primary significance of our findings lies in the systematic experimental exploration of a relatively understudied class of granular materials, namely those built from Platonic polypods.
As demonstrated by various authors, these materials possess intriguing and occasionally counter intuitive mechanical properties.
This suggests their potential applicability across diverse real-world contexts, such as architectural and civil engineering applications \cite{Dierichs2013,Dierichs2017}.
They offer stability without necessitating cementing materials and enable multiple assembly and disassembly cycles.
Furthermore, continued investigation into the microstructure of these materials, leveraging advanced imaging techniques, remains crucial.
We anticipate expanding our research efforts in this direction.

\begin{acknowledgments}
This project is supported by the LabEx NUMEV within the I-Site MUSE (ANR 2011-LABX-076).
The authors also acknowledge financial support from ANR MICROGRAM (ANR-20-CE92-0009).
We would like to express our gratitude to Gille Camp for its technical assistance in setting up the experimental device.
Additionally, we extend our thanks to Arnaud Regazzi, Benjamin Gallard, Sylvain Buonomo and the cfo-outillage company for their assistance in particle preparation.
\end{acknowledgments}

\bibliographystyle{unsrt}
\bibliography{references.bib}

\end{document}